\documentclass[aps,prb,twocolumn,superscriptaddress,showpacs,preprintnumbers,floatfix]{revtex4}
\usepackage{graphicx}
\usepackage{amssymb}
\usepackage{amsfonts}
\usepackage{amsmath}
\usepackage{color}
\newcommand{\ang}{\textup{\AA}}
\newcommand{\Fe}[1]{{Fe$^{(#1)}$}}
\newcommand{\surf}[0]{{\mathcal A}}
\newcommand{\gammaGB}[0]{\gamma_{\rm GB}}
\newcommand{\deltaGB}[0]{\delta_{\rm GB}}
\newcommand{\sigmaS}[0]{\sigma_{\rm surf}}
\newcommand{\Wsep}[0]{W_{\rm sep}}

\newcommand{\Vat}[0]{V_{\rm at}}
\newcommand{\muat}[0]{\mu_{\rm at}}

\usepackage{graphpap}

\begin{document}
\date{\today}

\title{Stability and magnetic properties of grain boundaries in the inverse Heusler phase Fe$_2$CoGa and in bcc Fe}

\author{Daniel F. Urban}\email{daniel.urban@iwm.fraunhofer.de}
\affiliation{Fraunhofer Institute for Mechanics of Materials IWM, W\"ohlerstr. 11, 79108 Freiburg, Germany}
\affiliation{University of Freiburg, Freiburg Materials Research Center, Stefan-Meier-Str. 21, 79104 Freiburg, Germany}
\author{Wolfgang K\"orner}
\affiliation{Fraunhofer Institute for Mechanics of Materials IWM, W\"ohlerstr. 11, 79108 Freiburg, Germany}
\author{Christian Els\"asser}
\affiliation{Fraunhofer Institute for Mechanics of Materials IWM, W\"ohlerstr. 11, 79108 Freiburg, Germany}
\affiliation{University of Freiburg, Freiburg Materials Research Center, Stefan-Meier-Str. 21, 79104 Freiburg, Germany}

\begin{abstract}
We investigate grain boundaries (GBs) in the cubic inverse Heusler phase Fe$_2$CoGa by means of first principles calculations based on density functional theory. Besides the energetic stability, the analysis focuses on the magnetic properties of a set of 16 GB structures in this intermetallic phase. We determine the integrated excess magnetization across the GB and analyze it in terms of the projected local magnetic moments of the atoms and their local Voronoi volumes. The results are systematically compared to those of corresponding GBs in body-centered cubic (bcc) Fe. 
The studied GBs in Fe$_2$CoGa may have a considerably increased magnetization at the GB, up to more than twice as much as in bcc Fe, depending on the GB type, while geometrical quantities like GB widening or local GB excess volume distributions are similar for both phases. We explain this difference by the higher flexibility of the ternary Fe$_2$CoGa phase in compensating the disturbance of a crystal defect by structural relaxation. The GB structures therefore have a lower energy accompanied with increased local magnetic moments of the Co and half of the Fe atoms within a distance of a few $\ang$ around the GB plane. 
\end{abstract}

\maketitle

\section{Introduction}
\label{sec:introduction}

The search for new low-cost hard magnetic compounds recently revealed several promising routes. 
Low rare-earth content intermetallic phases like the ThMn$_{{\rm 12}}$ crystal structure have
attracted renewed experimental and theoretical
interests.\cite{zh14,zh15,Goll14,hi15,mi14,ha15,ha15b,ko16}
Completely rare-earth free 
Heusler phases are promising candidates for which
high Curie temperatures,\cite{Takeuchi2003} magnetization, and
energy products (BH)$_{{\rm max}}$ have been achieved.\cite{He2020}

Several iron-rich Heusler phases with high magnetic moments like Fe$_2$CoGa or Fe$_2$CoAl have been identified theoretically.\cite{Gil2009,Gil2010} 
Unfortunately, the majority of Heusler phases with a high magnetic moment crystallize in the regular or inverse cubic Heusler structure with vanishing magnetocrystalline anisotropy.\cite{Gasi2013}
However, a substantial intrinsic crystalline anisotropy with a defined easy axis is needed for a good hard magnetic material. 
One idea to overcome this limitation is to deform the cubic phase into a tetragonal, trigonal, or hexagonal form.\cite{Zhang2015} Various authors have predicted tetragonally distorted regular or inverse Heusler phases with high magnetocrystalline anisotropy by systematic screenings using density functional theory (DFT).\cite{Faleev2017,Mats2017,Herper2018}
Apart from considering the influence of doping on different sublattice sites also the incorporation of additional light interstitial elements was studied as means to stabilize a tetragonal crystal structure.\cite{Zhang2014,Gao2020}
For thin films, epitaxially induced mechanical strain is a possible way to increase the crystallographic and magnetocrystalline anisotropy.\cite{Pechan2005}

Another cause of breaking the cubic symmetry is the presence of extended defects like grain boundaries (GBs). The smaller the grain size in a polycrystalline microstructure, the more relevant the GBs become. 
The proportion of atoms at GBs can reach double-digit percentages in nanocrystalline materials.\cite{gl89,pa90}
Nanoparticles of Co$_2$FeGa were already successfully synthesized by Basit et al.\cite{ba09} 
and a slight magnetic hardening compared to the polycrystalline bulk material was reported.
 
Following this idea, we study GBs in the ferromagnetic Heusler alloy Fe$_2$CoGa in this paper. This compound has a magnetic moment of about 5.3$\mu_B$ per formula unit which is one 
of the highest of all regular and inverse Heusler phases studied so far.\cite{Gil2009,Gil2010,Faleev2017,Mats2017,Herper2018,dan10}
The influence of several types of Fe$_2$CoGa GBs on the mechanical and magnetic properties is investigated by means of atomistic simulations based on DFT. 
This study comprises sixteen stable GBs in the inverse Heusler phase Fe$_2$CoGa for which we have  determined the grain boundary energy $\gamma$, the work of separation $\Wsep$, the local magnetic moments, and their respective accumulated change across the boundary region. 
For comparison seven GBs in pure body centered cubic (bcc) Fe have been analyzed, too. Similarities as well as differences between the two material systems are described and discussed.
The understanding gained by this study points out possibilities for a directed improvement of the desired physical properties.
Furthermore, the information about the magnetic moment distribution and
the inter grain interactions may be useful for
micromagnetic calculations using finite-element methods that provide insight
in the demagnetization processes of nanocrystalline ferromagnetic materials.\cite{fi96}

The manuscript is organized as follows: Section~\ref{sec:theo} 
presents the atomistic structure models and we define the energetic and magnetic quantities used to characterize the different GBs.  Section \ref{sec:results} reports the results on structural, energetic, and magnetic properties of the GBs. A detailed discussion on the distribution of local magnetic moments in the vicinity of the GB planes is presented. 
Section ~\ref{sec:summary} summarizes our findings.

\section{Theoretical approach}
\label{sec:theo}

\subsection{The inverse Heusler phase Fe$_2$CoGa}

\begin{figure}[tbp]
	\includegraphics[width=\columnwidth]{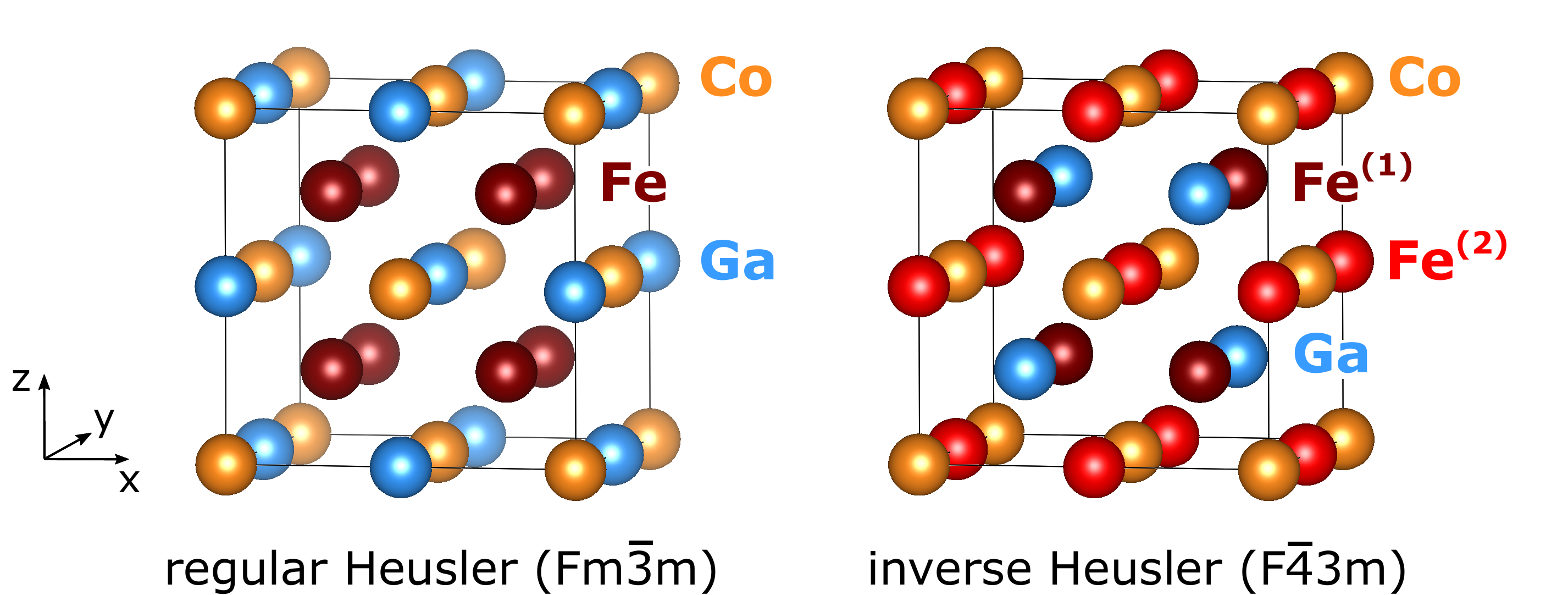}
\caption{
(Color online) Crystal structure of Fe$_2$CoGa in the regular (left) and inverse 
(right) full Heusler structure. Iron atoms are drawn in light and dark red which distinguishes between the two Fe sublattices of the inverse Heusler structure, while Co and Ga atoms are drawn in orange and blue, respectively.  \label{fig:bulk}}
\end{figure}

As a reference and starting point for the investigation of grain boundaries 
the conventional cubic unit cells of the full regular and inverse Heusler phases of Fe$_2$CoGa are shown in Fig.~\ref{fig:bulk}. Both crystal structures consist of four nested face-centered cubic (fcc) sublattices based at (0,0,0), ($\frac{1}{4}$, $\frac{1}{4}$, $\frac{1}{4}$), ($\frac{1}{2}$,$\frac{1}{2}$ ,$\frac{1}{2}$), and ($\frac{3}{4}$, $\frac{3}{4}$, $\frac{3}{4}$).

In the regular full Heusler phase (space group no.\ 225, Fm$\bar{3}$m) the Fe atoms are placed at Wyckoff positions 8c,  whereas the Co and Ga atoms occupy Wyckoff positions 4a (0,0,0) and 4b ($\frac{1}{2}$,$\frac{1}{2}$ ,$\frac{1}{2}$), respectively.

For the inverse Heusler phase (space group no.\ 216, F$\bar{4}$3m) 
two inequivalent Fe fcc sublattices are located at ($\frac{3}{4}$, $\frac{3}{4}$, $\frac{3}{4}$) and ($\frac{1}{2}$,$\frac{1}{2}$ ,$\frac{1}{2}$) (labeled \Fe1 and \Fe2 in the following) and Co and Ga atoms occupy the sublattices based at (0, 0, 0) and ($\frac{1}{4}$, $\frac{1}{4}$, $\frac{1}{4}$), respectively. Here, the corresponding Wyckoff positions are 4d (\Fe1), 4b (\Fe2), 4a (Co), and 4c (Ga). The different atomic neighborhood relationships for the \Fe1 and \Fe2 atoms has a strong influence on the local magnetic moments of the atoms, as summarized in Tab.\ \ref{tab:X}.

In the case of Fe$_2$CoGa, the inverse Heusler phase is energetically preferred compared to the regular Heusler structure.\cite{dan10}
Note that a substitution by Fe on all the lattice sites results in the genuine bcc structure of pure iron.

\begin{table}[]                               
\begin{tabular}{ccccc}                                                                                                
 \hline
 \hline
atom & nearest neighbors & 2. shell & 3. shell & $\mu/\mu_{\rm B}$ \\
\hline
\Fe1 & 4 \Fe2+4 Co & 6 Ga & 12 \Fe1 & 2.60\\
\Fe2 & 4 \Fe1+4 Ga & 6 Co & 12 \Fe2 & 1.86\\
Co     & 4 \Fe1+4 Ga & 6 \Fe2 & 12 Co & 1.09\\
Ga     & 4 \Fe2+4 Co & 6 \Fe1 & 12 Ga & -0.13\\
 \hline
 \hline
\end{tabular}
\caption{Overview of the atomic neighborhood relationships in the inverse Heuseler Fe$_2$CoGa crystal. The nearest neighbors are at a distance of $(\sqrt{3}/4)a_{\rm lat}$ while the second and third shell of atoms are at $a_{\rm lat}/2$ and $a_{\rm lat}/\sqrt{2}$, respectively. The different atomic environment result in different magnetic moments for the two iron sublattices. 
\label{tab:X} 
 }                                     
\end{table}

\subsection{GB models in inverse Heusler Fe$_2$CoGa}

\begin{figure*}
	\includegraphics[width=2\columnwidth]{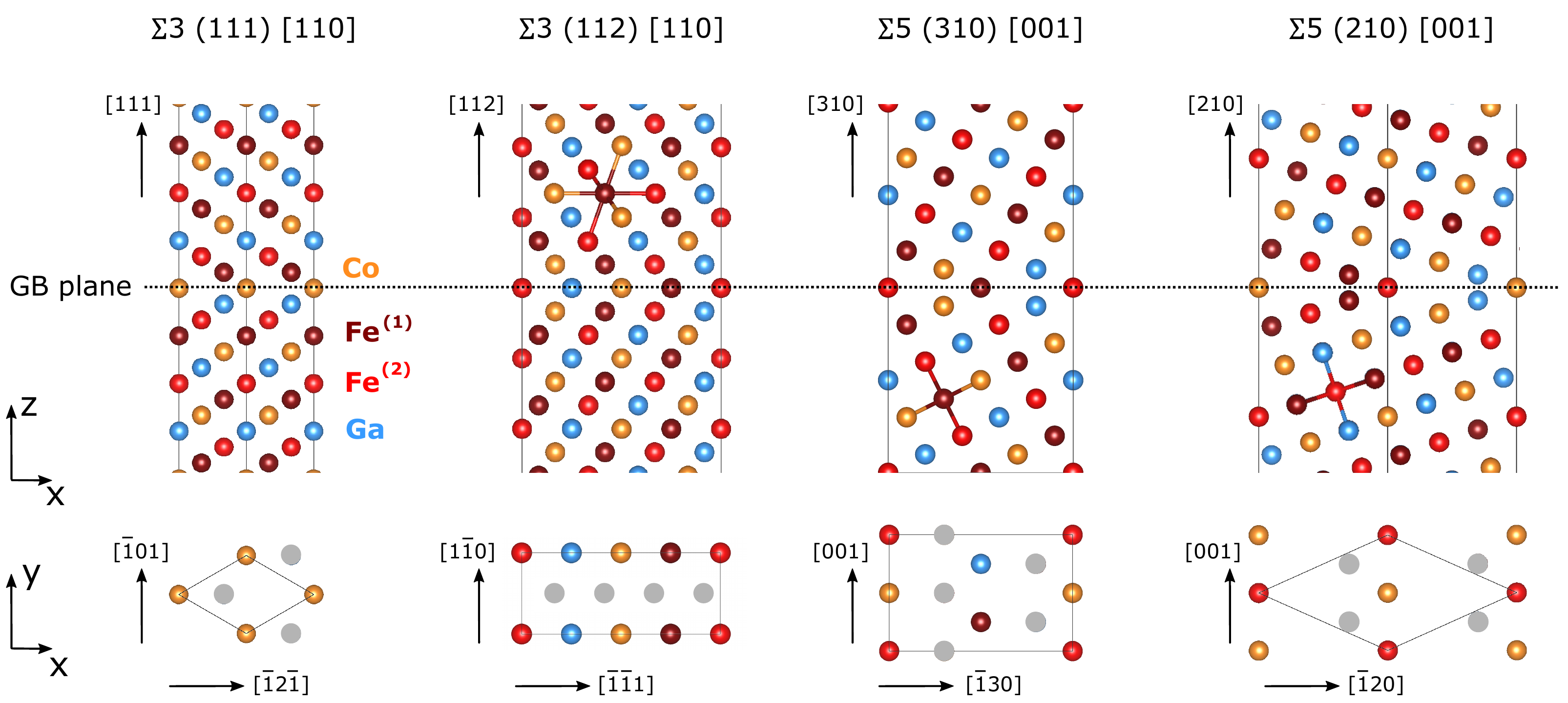}
\caption{(Color online)  Supercell models of the four prototypes of GBs studied in this work. For clarity, the mirror symmetric configuration (zero lateral translation) is shown in the upper part of the figure with the z-axis being perpendicular to the GB plane. The positions of the nearest neighbor atoms are exemplary shown for selected atoms by the drawn bonds.
The atomic configurations of the GB planes are shown in the lower part of the figure. Gray spheres mark the position of the atoms of the neighboring atom layers.
All the indicated lattice directions refer to the bulk crystal of the cubic inverse Heusler phase. 
\label{fig:GBsupercells}
}
\end{figure*}

For our study we have chosen four types of symmetric tilt grain boundaries (STGB), namely $\Sigma 3$ (111)[110], $\Sigma 3$ (112)[110], $\Sigma 5$ (210)[001], and $\Sigma 5$ (310)[001] as representatives of typical low energy tilt GBs in the inverse Heusler Fe$_2$CoGa.
Note that for simplicity of notation the tilt axis [110] for $\Sigma 3$
and [001] for $\Sigma 5$  will be omitted in the following. 

For each GB type we have constructed supercell models as visualized in Fig.~\ref{fig:GBsupercells} and compiled in Tab.~\ref{tab:sc:dim}. In all structure models we refer to the axis perpendicular to the GB plane as z axis, while the x and y axes span the GB plane. The tilt axis of the GB is along the y axis.

\begin{table}[]  
\begin{tabular}{ccccccc}                                                                                       
\hline
\hline
compound & GB    &  $N_{\rm at}$ & $L_z^{(0)}$ [$\mathring{A}$] & $\surf$ [$\mathring{A}^2$]   & $n_{\rm term.}$ & $n_{\rm trans.}$ \\
\hline  
Fe$_2$CoGa & $\Sigma 3$ (111)   &48& 39.70 & 14.22 & 4 & 6 \\ 
Fe$_2$CoGa & $\Sigma 3$ (112)   &96& 28.08 & 40.22 & 1 & 12\\ 
Fe$_2$CoGa & $\Sigma 5$ (210)   &80& 25.63 & 36.72 & 2 & 8 \\ 
Fe$_2$CoGa & $\Sigma 5$ (310)   &80& 18.12 & 51.93 & 1 & 12\\ 
\hline
bcc Fe     & $\Sigma 3$ (111)   &24& 19.63 & 13.90 & 1 & 1 \\
bcc Fe     & $\Sigma 3$ (112)   &24& 27.75 &  9.83 & 1 & 2 \\ 
bcc Fe     & $\Sigma 5$ (210)   &40& 26.10 & 17.95 & 1 & 2 \\ 
bcc Fe     & $\Sigma 5$ (310)   &40& 36.73 & 12.69 & 1 & 2 \\ 
\hline
\hline
\end{tabular}                                                                                
\caption{Geometric characteristics of the pristine supercell models for GBs in inverse Heusler Fe$_2$CoGa and in bcc Fe. Listed are the number of atoms $N_{\rm at}$ in the supercell model, the supercell dimension $L_z^{(0)}$ perpendicular to the GB plane (for zero excess widening) and the GB area $\surf$. The last two columns list the number $n_{\rm term.}$ of possible nonequivalent GB planes (terminations) and the number $n_{\rm trans.}$ of considered relative rigid body translations in xy plane of the two crystal blocks forming the GB.
\label{tab:sc:dim}}       
\end{table}

Several relative lateral translations of the two grains parallel to the GB plane were considered and structurally optimized for each GB type (see Tab.\ \ref{tab:sc:dim} for an overview). For the $\Sigma 3$ (111) and  $\Sigma 5$ (210) GB types, different permutations of the sequence of atomic layers are possible which result in inequivalent GB configurations. After the structural relaxation, only a few  nonequivalent low energy configurations remained for each GB type, which were then analyzed further in detail.

For the structural optimization of the GB structure models the GB area was always kept fixed according to the lattice parameters of the bulk structure. The supercell dimension perpendicular to the interface was varied systematically in steps of 0.1 {\AA}. For every separation of the two grains a structural relaxation of the atom positions was carried out. The optimal geometry and grain separation were extracted by fitting the obtained relationship of total energy versus lateral dimension. The GB excess widening is thus obtained as
\begin{equation}
\deltaGB = \frac{c_{\rm GB} - c_{\rm bulk}}{2},
\end{equation}
where $c_{\rm GB}$ and $c_{\rm bulk}$ are the optimized supercell dimensions in z-direction of the GB structure model and the corresponding bulk structure, respectively.

\subsection{GB structures in bcc Fe}

\begin{figure}
	\includegraphics[width=\columnwidth]{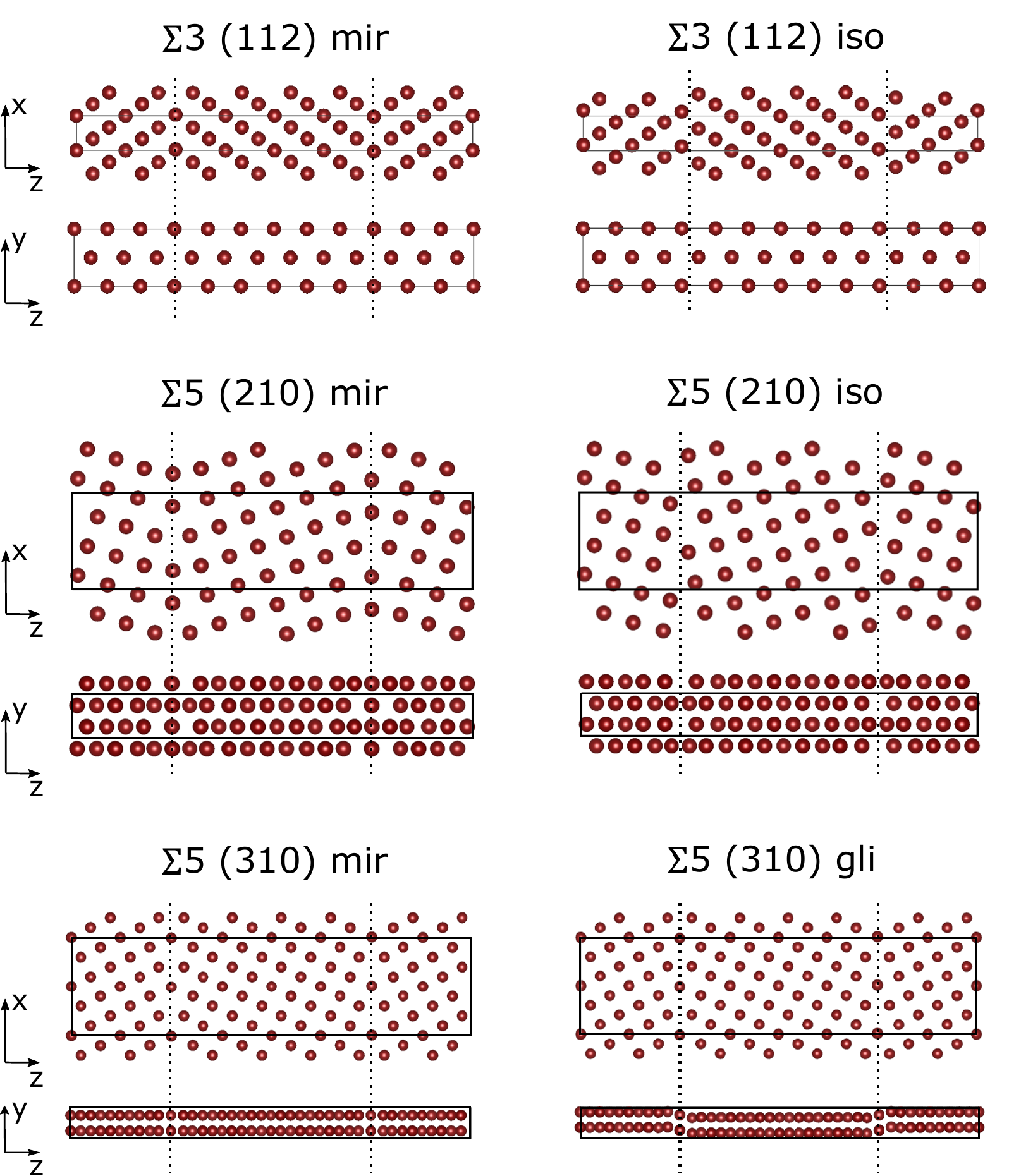}
\caption{(Color online)  Supercell models of GBs in bcc Fe. The GB planes are given by symmetry planes and are indicated by dotted lines.
\label{fig:FeGB}
}
\end{figure}

We considered seven structural models of GBs in bcc Fe (cf.\ Tab.~\ref{tab:sc:dim} and Fig.~\ref{fig:FeGB})
Note that due to the higher symmetry in pure bcc Fe, the respective supercell models can be reduced in size compared to the models for the inverse Heusler phase Fe$_2$CoGa. 

For both, the $\Sigma$3 (112) and the $\Sigma$5 (210) we constructed two realizations, namely GBs of mirror and isosceles type.\cite{Mrovec2003} In the isosceles structure, the two grains are translated perpendicular to the tilt axis and parallel to the GB plane so that the interatomic bonds across the GB plane form a pattern of isosceles triangles.
For the $\Sigma$5 (310) there exists a mirror symmetric form (see e.g.\ Ref.\ \onlinecite{ca08}) $\Sigma$5 (310)$_{{\rm mir}}$ and a $\Sigma$5 (310)$_{{\rm gli}}$ which includes a lateral translation of in y direction along the tilt axis between the two grains corresponding to a glide reflection.\cite{Ochs2000}
Finally, the $\Sigma$3 (111) is mirror symmetric.

\subsection{Energetic characterization of grains boundaries}
\label{sec:physical:quantities}

For each GB we computed the characteristic energetic properties: grain boundary energy $\gammaGB$, surface energy $\sigmaS$, and work of separation 
$\Wsep$. These quantities are obtained from total energy differences,
\begin{eqnarray}
  \gammaGB  &=& \frac {E_{\text{GB}}- E_{\text{bulk}}}{2 \surf},\\
  \Wsep     &=& \frac {E_{\text{GB}}-2E_{\text{FS}}} {2 \surf}, \\
  \sigmaS   &=& \frac {E_{\text{FS}}- E_{\text{bulk}}/2} {2 \surf},
\end{eqnarray}
where $E_{\text{GB}}$ is the total energy of the GB supercell, $E_{\text{bulk}}$ 
is the total energy of the supercell of the perfect single crystal structure of same size, and $\surf$ is the area of the GB plane.
$E_{\text{FS}}$ is the total energy of the supercell of half size, where the upper grain has been removed from the GB supercell and replaced by vacuum. Note that the three quantities $\gammaGB$, $\sigmaS$, and $\Wsep$ are related by 
\begin{eqnarray}
\label{eq:relation}
   \Wsep+\gammaGB-2\sigmaS&=&0.
\end{eqnarray}
When separating the two grains forming the $\Sigma 3$ (111) or $\Sigma 5$ (210) GBs in the inverse Heusler structure, two nonequivalent surfaces are created. In theses cases, $\sigmaS$ represents their mean value.

\subsection{Magnetic characterization of grain boundaries}
\label{sec:def:DeltaM}
We monitor the change of magnetic moment due to the presence of a GB by 
\begin{equation}
\label{eq:xiM}
   \xi_M= \frac{M_{\text{GB}}- M_{\text{bulk}}} {2 \surf},
\end{equation}
where $M_{\text{GB}}$ and $M_{\text{bulk}}$ are the total magnetic moments of the GB supercell
and of the corresponding bulk supercell, respectively. 
The absolute change $\Delta M = M_{\text{GB}}- M_{\text{bulk}}$ of the total magnetic moment due to the presence of the GB is typically of the order of a few Bohr magnetons $\mu_B$.

For a subset of the GB models, namely the $\Sigma 3$ (112) and $\Sigma 5$ (210) GBs of the inverse Heusler phase, several supercell models contain two chemically inequivalent GBs.
In principle this would require separate analyses of the changes in the magnetic moments for the two GBs within the supercell. However, both results are found to be very similar and for simplicity we will always list and discuss the mean values of $\xi_M$ for these cases.

While the energetic quantities converge rather quickly with increasing supercell dimension perpendicular to the GB, a convergence of the magnetic moments of the individual atoms in the intermediate bulk-like region requires substantially larger cells. 
For example, the supercell models of a $\Sigma$3 (112) GB with 48 atoms in the inverse Heusler phase have essentially one ``bulk'' layer of atoms in the center between the two GBs, with atoms having local magnetic moments comparable to the bulk values. On the other hand, for the larger supercells with 96 atoms, there is a region of approximately six ``bulk'' layers for which the magnetic moments correspond well to the values obtained for the single crystal.

Apart from the excess magnetization per GB area $\xi_M$ at structural equilibrium we analyze the sensitivity of this quantity with respect to an applied elastic strain in z direction, i.e. perpendicular to the GB plane. We therefore evaluate
\begin{equation}
\frac{\partial\xi_M}{\partial\epsilon_z}=\frac{1}{2\surf}\left(
c_{\text{GB}}\frac{\partial M_{\text{GB}}}{\partial c} - 
c_{\text{bulk}}\frac{\partial M_{\text{bulk}}}{\partial c}
\right)
\end{equation} 
where the derivatives ${\partial M}/\partial c$ are obtained from linear fits to the data obtained for small variations in the c lattice parameter around its equilibrium value.

\subsection{Computational setup}    
\label{sec:vasp}

All DFT calculations were carried out using the projector augmented-wave (PAW) method\cite{bl94} as implemented in the VASP code.\cite{kr96,kr99} The PBE-GGA\cite{Perdew1996} was used for the exchange-correlation functional, and PAW pseudopotentials with 14, 9 and 3 valence electrons were used for Fe (3p$^6$3d$^7$4s$^1$), Co (3d$^8$ 4s$^1$), and Ga(4s$^2$3p$^1$), respectively.       
All calculations were performed with spin-polarization and with a plane-wave cutoff energy of 600 eV. The Brillouin-zone integrals were sampled by $8\times 8\times 8$ Monkhorst-Pack k-point grids for the 16-atom unit cells, and a Gaussian broadening of 0.05 eV.
Accordingly, the specific k-point grids for the various Fe$_2$CoGa and bcc Fe supercell models containing GBs or surfaces were adjusted to approximately the same k-point density.
Interatomic forces were relaxed using the Broyden-Fletcher-Goldfarb-Shanno (BFGS) algorithm.

\section{Results}
\label{sec:results}

\subsection{Bulk phases}
We have calculated the equilibrium cohesive, structural, and magnetic properties of the Fe$_2$CoGa regular and inverse Heusler phases.  
Our results are summarized in Tab.\ \ref{tab:results:bulk} and are in good agreement with reported values of experimental and other theoretical work.
The inverse Fe$_2$CoGa Heusler phase has its energetic minimum for the cubic case ($c/a = 1$) and its formation energy is 0.40 eV/f.u. lower than that of the regular Fe$_2$CoGa Heusler phase.

\begin{table}[]  
\begin{tabular}{llccc}  
\hline
\hline
structure    & property & this work & other DFT & Exp. \\   
\hline 
reg. Fe$_2$CoGa& $a_{\rm lat}$ [\AA]        & 5.762  & 5.774\cite{dan10}& \\
               & $\mu$ [$\mu_{\rm B}$/f.u.] & 6.09   & 6.08\cite{dan10} & \\
\hline
inv. Fe$_2$CoGa& $a_{\rm lat}$ [\AA]        & 5.731  & 5.736\cite{dan10},5.727\cite{Gil2010}  & 5.71\cite{jag78} \\
               & $\mu$ [$\mu_{\rm B}$/f.u.] & 5.42   & 5.29\cite{dan10}, 5.27\cite{Gil2010}   & \\
\hline
bcc Fe         & $a_{\rm lat}$ [\AA]        & 2.833 & 2.84\cite{Paxton2010} & 2.866\cite{ho67} \\
               & $\mu$ [$\mu_{\rm B}$/f.u.] & 2.19  & 2.2\cite{Paxton2010}& 2.2\cite{ii13} \\
\hline
\hline
\end{tabular}                                                                                
\caption{Equilibrium lattice constants $a_{\rm lat}$ and magnetic moments $\mu$ for Fe$_2$CoGa in the regular and inverse Heusler phase and for bcc iron. The comparison of our results with experimentally measured values as well as with other theoretical work shows a good agreement.
\label{tab:results:bulk}
}       
\end{table}

\subsection{Stable low-energy GB structures in Fe$_2$CoGa}

For the $\Sigma 3$ (111) GB only one of all the initially different translation states is found to be stable, namely the mirror symmetric configuration with zero lateral translation. There are four possible realizations that differ by the atom species populating the mirror plane which we accordingly denote by 
$\Sigma 3$ (111)$_{X}$ with X = \Fe1, \Fe2, Co, or Ga.

For the $\Sigma 3$ (112) GB altogether four stable low-energy configurations were found after relaxation. Apart from the mirror symmetric configuration $\Sigma 3$ (112)$_{\rm mir}$ (with zero lateral translation) three further low-energy GBs were identified with translation dx = 0.25, 0.5, and 0.75 (given in fractions of the extension in x direction) which we label $\Sigma 3$ (112)$_{dx}$.
All these supercell models contain two chemically nonequivalent GBs which is unavoidable for the considered crystal structure. All GB properties are thus averages over the respective two GBs within the structure model. However, from the respective surface energies one can see that the chemical termination has very little influence on the absolute values.

The $\Sigma 5$ (210) GB type provides a rich set of possible GB configurations. Starting from the mirror symmetric configuration we have analyzed all $4\times2$ combinations of lateral translations $dx = 0$, 0.1, 0.5, 0.6 and $dy = 0$, 0.25. Moreover, for each GB configuration there are two possible choices of defining 
the GB plane, namely the two terminations \Fe2/Co and \Fe1/Ga. A total of four stable low energy configurations were determined by structural optimization. For one of these configurations, labeled $\Sigma 5$ (210)$_{\rm mir}$, the GB is a true mirror plane, while for a second one the atom positions to both sides of the GB are close to mirror symmetry, but the atom types differ. We label this pseudo-mirror GB model $\Sigma 5$ (210)$_{\rm psmir}$. Furthermore, we found a  
$\Sigma 5$ (210)$_{\rm iso}$ of isosceles-type and finally a $\Sigma 5$ (210)$_{\rm mix}$ with GBs of mixed type.

For the $\Sigma 5$ (310) GB the structural optimization yields
four stable low energy configurations. Apart from  the mirror symmetric configuration they have relative lateral translation of the two grains along the tilt axis by dy = 0.25, 0.5, and 0.75 (given in fractions of the supercell extension in y direction) We label them by $\Sigma 5$ (310)$_{\rm mir}$ and $\Sigma 5$ (310)$_{dy}$, respectively.

\begin{table*}[]                                                           
\begin{tabular}{llcccccccc}                                                                                                
\hline
\hline
	  &	GB    
	  & $\gammaGB$
	  & $\Wsep$ 
	  & $\sigmaS$ 
	  & $\gammaGB/\sigmaS$
	  & $\xi_M$   
	  & $\frac{\partial\xi_M}{\partial\epsilon_z}$ 
	  & $\deltaGB$ 
	  & $\gammaGB$  (Literature)
	  \\
	  &    
	  & $[{J}/{m^2}]$
	  & $[{J}/{m^2}]$
	  & $[{J}/{m^2}]$
	  & 
	  & $[{\mu_B}/{nm^2}]$  
	  & $[{\mu_B}/{\ang^2}]$ 
	  & [\AA]
	  & $[{J}/{m^2}]$ 
	  \\
\hline                                                                                                             
Fe$_2$CoGa                        
&$\Sigma 3$ (112)$_{{\rm mir}}$	& 0.35 & 3.92 &	2.13 &	0.16 &	5.13 & 0.52 &	0.149 &\\
&$\Sigma 3$ (112)$_{0.25}$	    & 0.33 & 3.93 &	2.13 &	0.16 &	5.12 & 0.64 &	0.136 &\\
&$\Sigma 3$ (112)$_{0.5}$	    & 0.32 & 3.95 &	2.13 &	0.15 &	4.95 & 0.59 &	0.131 &\\
&$\Sigma 3$ (112)$_{0.75}$	    & 0.34 & 3.93 &	2.13 &	0.16 &	6.44 & 0.66 &	0.145 &\\
\hline
&$\Sigma 3$ (111)$_{{Fe^{(1)}}}$& 0.97 & 3.69 &	2.33 &	0.42 &	6.61 & 1.59 &	0.300 &\\
&$\Sigma 3$ (111)$_{{Fe^{(2)}}}$& 1.26 & 3.24 &	2.25 &	0.56 &	8.37 & 1.13 &	0.338 &\\
&$\Sigma 3$ (111)$_{Co}$		& 1.19 & 3.45 &	2.32 &	0.51 &	8.70 & 1.61 &	0.245 &\\
&$\Sigma 3$ (111)$_{Ga}$		& 0.96 & 3.48 &	2.22 &	0.43 &	7.38 & 1.28 &	0.296 &\\
\hline
&$\Sigma 5$ (310)$_{{\rm mir}}$ & 1.18 & 3.08 &	2.13 &	0.55 &	7.13 & 0.32 &	0.440 &\\
&$\Sigma 5$ (310)$_{0.25}$	    & 1.18 & 3.08 &	2.13 &	0.55 &	7.25 & 0.11 &	0.443 &\\
&$\Sigma 5$ (310)$_{0.5}$	    & 1.05 & 3.21 &	2.13 &	0.49 &	6.98 & 0.11 &	0.431 &\\
&$\Sigma 5$ (310)$_{0.75}$      & 1.18 & 3.08 &	2.13 &	0.55 &	7.23 & 0.14 &	0.442 &\\
\hline
&$\Sigma 5$ (210)$_{{\rm mir}}$	 & 1.07 & 3.15 & 2.11 &	0.51 &	8.88 & 0.12 &	0.398 &\\
&$\Sigma 5$ (210)$_{{\rm psmir}}$& 1.08	& 3.14 & 2.11 &	0.51 &	10.06& 0.19 &	0.423 &\\
&$\Sigma 5$ (210)$_{{\rm iso}}$  & 1.12	& 3.10 & 2.11 &	0.53 &	8.73 & 0.16 &	0.438 &\\
&$\Sigma 5$ (210)$_{{\rm mix}}$	 & 1.16	& 3.06 & 2.11 &	0.55 &	9.64 & 0.14 &	0.430 &\\
\hline
\hline  
bcc-Fe
&$\Sigma 3$ (112)$_{{\rm mir}}$ & 0.45 & 4.83 & 2.64 & 0.17 & 4.71 & 0.49 & 0.123 & 0.34\cite{ga09}, 0.43\cite{Bhat2014}, 0.47\cite{bh13,du11}\\
&$\Sigma 3$ (112)$_{{\rm iso}}$ & 0.42 & 4.86 & 2.64 & 0.16 & 5.33 & 0.56 & 0.117 & 0.45\cite{Wang2018}\\
&$\Sigma 3$ (111)				& 1.60 & 3.90 & 2.75 & 0.58 & 3.70 & 1.50 & 0.306 & 1.52\cite{ga09}, 1.57\cite{wa10}, 1.61\cite{Bhat2013}\\						
&$\Sigma 5$ (310)$_{{\rm mir}}$	& 1.58 & 3.64 & 2.61 & 0.60 & 3.44 & 1.38 & 0.315 & 1.53\cite{du11}, 1.55\cite{Tahir2014}, 1.57\cite{Wang2018}, 1.63\cite{ca08}\\
&$\Sigma 5$ (310)$_{{\rm gli}}$	& 1.54 & 3.68 & 2.61 & 0.59 & 2.49 & 0.76 & 0.313  & \\
&$\Sigma 5$ (210)$_{{\rm mir}}$	& 1.62 & 3.58 & 2.60 & 0.62 & 8.37 & 0.91 & 0.368  & \\
&$\Sigma 5$ (210)$_{{\rm iso}}$	& 1.70 & 3.50 & 2.60 & 0.65 & 12.52 & 0.95 & 0.411 & 1.64\cite{Wang2018}, 2.00\cite{wa10}\\
\hline
\hline
\end{tabular}                                                                                
\caption{Geometric, energetic and magnetic characteristics of the model GBs in inverse Heusler Fe$_2$CoGa and bcc Fe. Listed are the grain boundary energy $\gammaGB$, work of separation $\Wsep$, surface energy $\sigmaS$, GB excess magnetization $\xi_M$, the strain derivative of the latter 
$\partial\xi_M/\partial\epsilon_z$, and the GB excess widening $\deltaGB$. For comparison, the last column lists $\gammaGB$ values of bcc Fe GBs obtained in DFT studies by other authors.
  \label{tab:results:GB}}       
\end{table*}

\subsection{Energetic properties of GBs}

\begin{figure}
	\includegraphics[clip, trim=1.cm 0.1cm 0.1cm 1.5cm,width=\columnwidth]{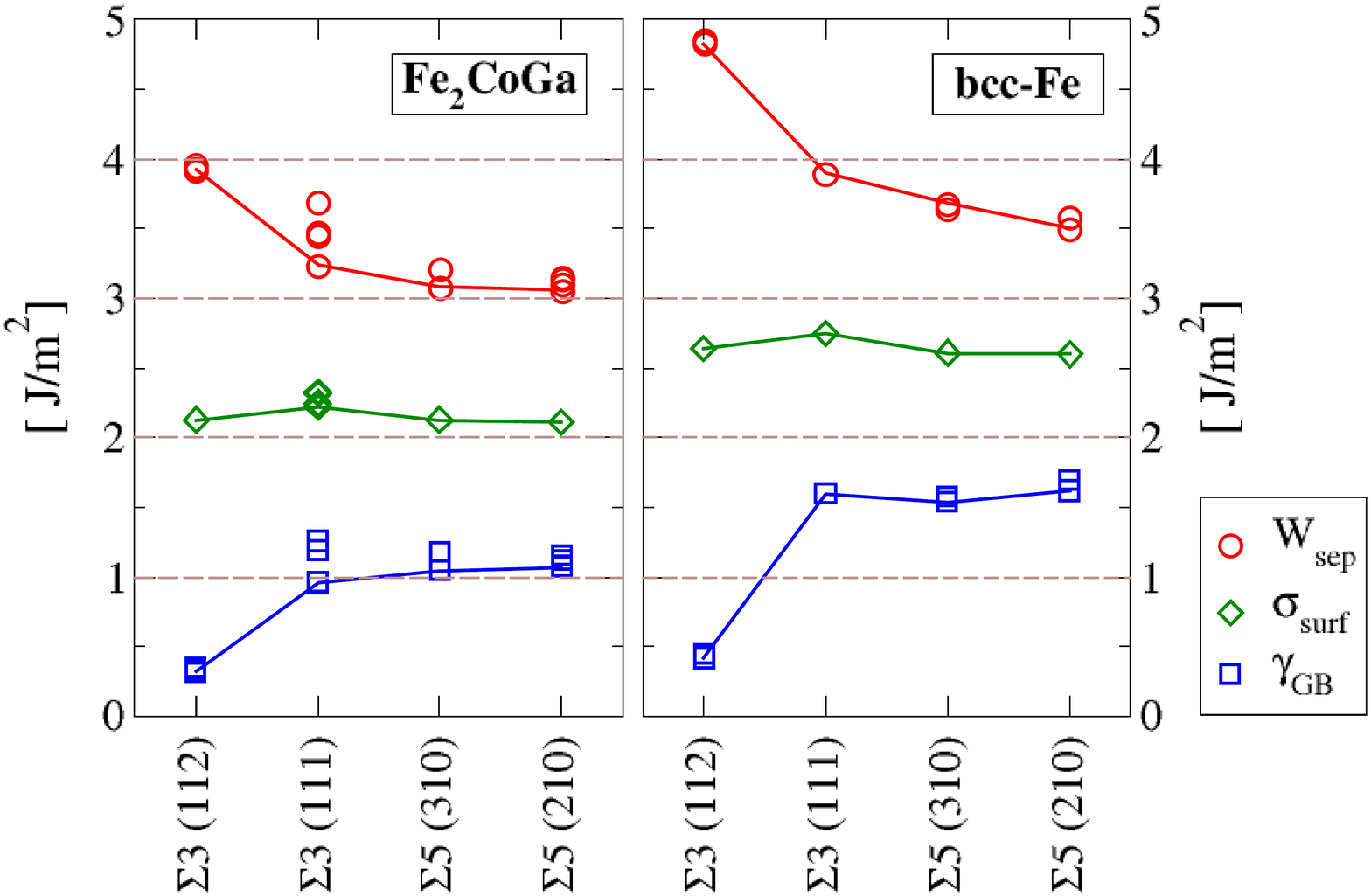}
\caption{(Color online) GB energy $\gammaGB$, work of separation $\Wsep$ and surface energy $\sigmaS$ of low energy GBs in the inverse Heusler Fe$_2$CoGa (left panel) and in bcc Fe (right panel). The lines connect the lowest energy values of each GB type and serve as guide for the eye.
\label{fig:GB_energetics}
}
\end{figure}

Our results for the GB energy $\gammaGB$, work of separation $\Wsep$ and  surface energy $\sigmaS$ of low-energy GBs in Fe$_2$CoGa are plotted and compared to the respective values of GBs in bcc Fe in Fig.~\ref{fig:GB_energetics}. The numerical values are compiled in Tab.~\ref{tab:results:GB}.

The formation of a GB is always connected with a cost in energy. The different realizations of the $\Sigma 3$ (112) GB in Fe$_2$CoGa have clearly the lowest GB energies which are about 25--35\%  of the energies of the other GBs. 
This is understandable from the fact that the atomic arrangement at the GB interface of $\Sigma 3$ (112) is very similar to the perfect bulk structure.

Interestingly, our values for the $\gammaGB$
of $\Sigma 3$ (111), $\Sigma 5$ (210) and $\Sigma 5$ (310)
in the inverse Heusler phase are considerably lower than in bcc Fe which means that they are relatively more likely to form. 
The reason for the higher $\gammaGB$ values in bcc Fe may lie in the lower ability
to compensate for the local structural perturbations at the GBs. In the inverse Heusler phase the three chemical elements provide a higher variability 
which enables more stable structural units at the GB.

The energies of the considered surfaces in Fe$_2$CoGa are also systematically lower by $\sim$ 0.5 J/m$^2$ that those of the respective bcc Fe surfaces. In both systems, $\sigmaS$ varies only weakly between the different surface orientations (111), (112), (210), and (310) which indicates the efficient electronic screening of the two metals.

The work of separation shows an opposite behavior compared to $\gammaGB$: The most stable $\Sigma 3$ (112) GBs (with the lowest $\gammaGB$ value) yield the largest $\Wsep$. This directly follows from Eq.\ (\ref{eq:relation}) in combination with the almost constant $\sigmaS$. For GBs in bcc Fe, $\Wsep$ is increased by 0.5--1 J/m$^2$ compared to F$_2$CoGa, as a consequence of the systematically larger surface energies.

Our results confirm the widespread heuristic relation that the GB energy $\gammaGB$ generally is about 1/2 to 2/3 of the surface energy values $\sigmaS$.\cite{Rohrer2011}
As can be seen from the values in Tab.~\ref{tab:results:GB} this correlation holds except for the $\Sigma 3$ (112) GBs. This is due to their significantly lower  
$\gammaGB$ values of the $\Sigma 3$ (112) while their surface formation energies $\sigmaS$ are comparable to those of the other GB types.

We find an overall good agreement of our computed GB energies in bcc Fe compared to other theoretical DFT studies reported in literature (see last column in Tab.\ \ref{tab:results:GB}).
Only for the $\Sigma 5$ (210)$_{\rm iso}$ GB in bcc Fe our $\gammaGB$ value is about 15\% smaller than reported by
Wachowiez et al.\cite{wa10}. This may be connected to the in comparison lower plane wave cutoff (350eV) taken by the authors. Test calculations with a lower plane wave cutoff and k-mesh yield higher $\gammaGB$ values.

\subsection{GB excess magnetization $\xi_M$}

\begin{figure}
	\includegraphics[clip, trim=2.cm 17.3cm 10.2cm 2.6cm,width=\columnwidth]{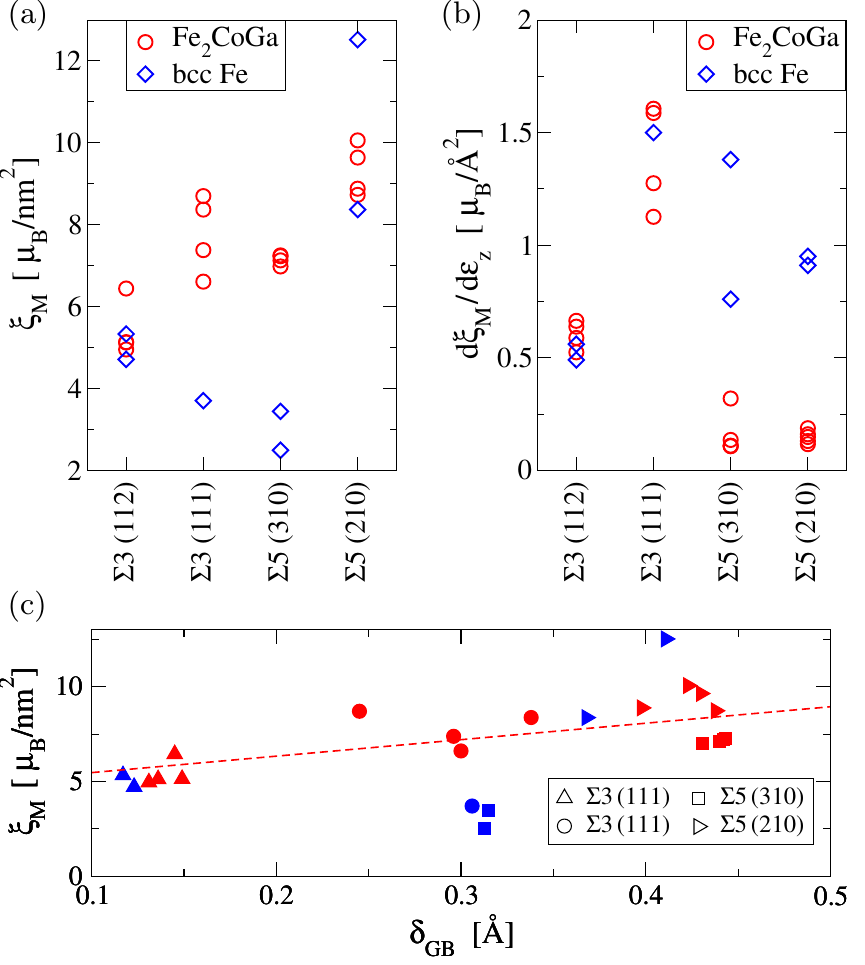}
\caption{(Color online) (a) Integrated GB excess magnetisation $\xi_M$  and (b) its sensitivity to a strain $\epsilon_z$ applied perpendicular to the GB, for low energy GBs in inverse Heusler Fe$_2$CoGa (red circles) and in bcc Fe (blue diamond symbols). (c) Correlation between  $\xi_M$ and the GB widening $\deltaGB$.  \label{fig:GB_magn_prop}
}
\end{figure}

A GB affects the total magnetization of the considered structure model. We quantify this change by the excess magnetization $\xi_M$, as defined in Eq.\ (\ref{eq:xiM}).
For all the systems we investigated, the integrated total magnetic moment always increases due to the presence of a GB, as can be seen from the $\xi_M$ values compiled in Tab.\ \ref{tab:results:GB} and  visualized in Fig.\ \ref{fig:GB_magn_prop} (a). The sensitivity of the magnetization with respect to strain (and thereby a change in local volume), expressed by the derivative $\partial\xi_M/\partial\epsilon_z$, is plotted in Fig.\ \ref{fig:GB_magn_prop} (b).

In comparison, the smallest increase in magnetization for the inverse Heusler phase GBs is found for the 
\mbox{$\Sigma 3$ (112)}, which in addition is very similar to the $\xi_M$ of the respective bcc Fe GBs. Moreover, the sensitivity to strain of this GB type is almost identical for both phases. This originates in the structural similarity and the comparably small deviation from bulk structures of the \mbox{$\Sigma 3$ (112)} GBs.

For the other three GB types, there are substantial differences between Fe$_2$CoGa and bcc Fe.
The $\Sigma 3$ (111) GBs in the inverse Heusler phase have a wide range of $\xi_M=6.6$--$8.7\; \mu_B/$nm$^2$, depending on the chemical elements at and adjacent to the GB plane. This enhancement is considerably larger than the $\xi_M=3.7\; \mu_B/$nm$^2$ observed for the $\Sigma 3$ (111) GB in bcc Fe. However, the sensitivity to strain is rather similar for both phases and among the largest values within the set of studied GB structures.

For the $\Sigma 5$ (310) GB the difference in $\xi_M$ between the two phases is even larger. Here, in the inverse Heusler phase, the atomic planes parallel to the GB contain all four elements, which leads to a high flexibility under structural relaxation. As a consequence, the $\xi_M$ values are found to be very similar for the four considered GB models and more than twice as large as the $\xi_M$ in bcc Fe. This increase in magnetization can mainly be attributed to increased local magnetic moments of Co and \Fe2 atoms at the GB, as will be shown below.  
Also the $\Sigma 5$ (210) GBs have a considerable spread in $\xi_M$ for both phases. This GB type results in the highest $\xi_M$ of the four types considered in this study. The $\Sigma 5$(210)$_{\rm iso}$ in bcc Fe has the overall largest GB excess magnetization. We discuss the origin thereof, especially in comparison with the mirror symmetric GB, in the following Sec.\ \ref{sec:local_moments}.
Interestingly, both the $\Sigma 5$ (210) and $\Sigma 5$ (310) GBs in the inverse Heusler phase are comparably insensitive to strain, i.e. small changes of the GB excess widening $\deltaGB$. In comparison, the respective GBs in bcc Fe have considerably larger $\partial\xi_M/\partial\epsilon_z$ values, cf.\ Fig.\ \ref{fig:GB_magn_prop} (b).

In Fig.\ \ref{fig:GB_magn_prop} (c) we plot the excess magnetization $\xi_M$ as function of the excess  widening $\deltaGB$. Although there is a slight average trend to an increase of $\xi_M$ with increasing  $\deltaGB$, no obvious correlation can be stated. There is a large scatter of the data within a given GB type, which reflects the importance of the local atomic neighborhood in the vicinity of the GB plane on the local atomic magnetic moments and therefore on the the total magnetization. This aspect will be investigated in detail in the following section.

\subsection{Local magnetic moments}    
\label{sec:local_moments}

\begin{figure}
\begin{center}
		\includegraphics[clip, trim=2.cm 18.2cm 10.2cm 1.7cm,width=\columnwidth]{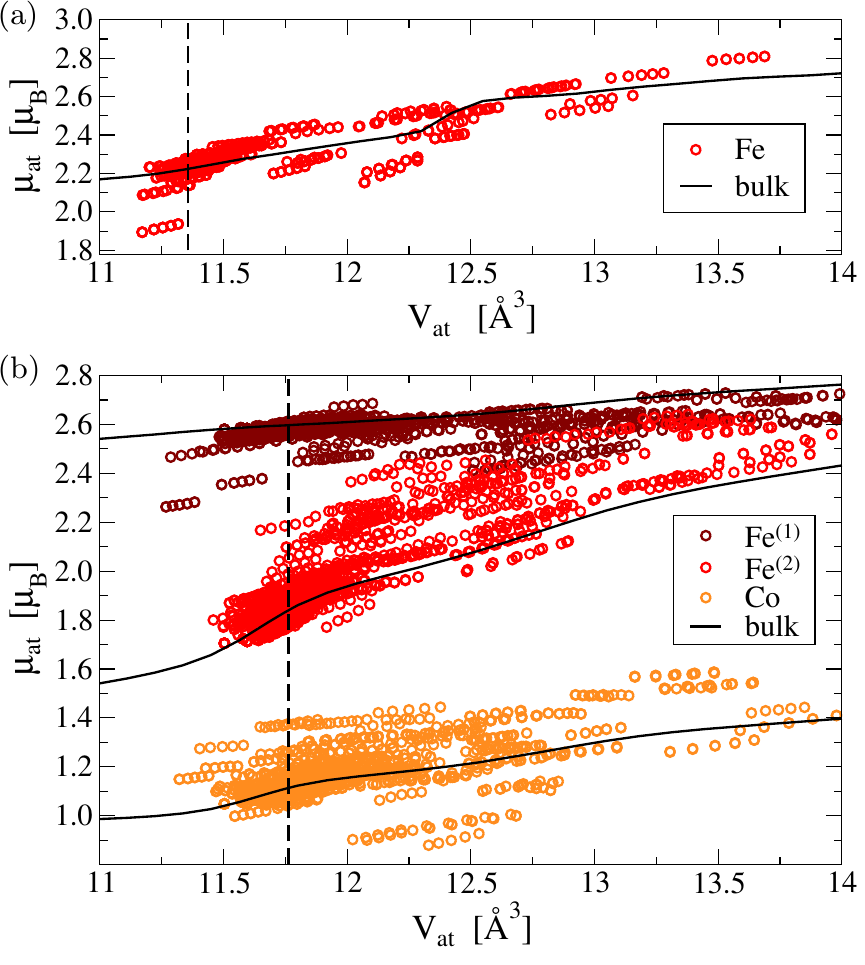}
 	\end{center} 
\caption{(Color online) Local magnetization $\muat$ as function of Voronoi volume $\Vat$ for all magnetic atoms in the studied GBs of (a) bcc Fe and (b) the inverse Heusler phase. Vertical dashed lines indicate the volume per atom of the equilibrium bulk structure. 
The local magnetic moments in the bulk phase as function of volume are shown as black lines for comparison. \label{fig:vol_vs_locmag} }
\end{figure}

An increase of the local magnetization at the GB
is often attributed to an increased local volume\cite{ca08,wa10,bh13}
which can be quantified, e.g., by a Voronoi cell volume for each atom. 
However, in bcc Fe the number of next neighbors and the respective bond 
lengths can vary and influence the absolute values of the local moment, too. 
Figure \ \ref{fig:vol_vs_locmag} (a) displays the distribution of local magnetic 
moments $\muat$ that we got for the bcc Fe GBs and their correlation with the Voronoi volume $\Vat$ of the atoms. For comparison, the solid line marks the result for an isotropic variation of volume for a bulk single crystal and the vertical dashed line indicates the equilibrium volume. Most of the atoms have a magnetic moment and a Voronoi volume close to the bulk values. Some atoms, especially those directly at the GB plane, may have a significantly increased $\muat$ up to 2.8 $\mu_B$ and a 20\% larger Voronoi volume.  

In the inverse Fe$_2$CoGa Heusler phase a fourth aspect, namely the chemical variation
of the neighborhood, strongly influences the local magnetic moments of Fe and Co.
The effect is already very pronounced in the bulk phase where the atoms \Fe1 and \Fe2
have exactly the same Voronoi volume but strongly deviating moments $\mu_{Fe1}$ = 2.60 $\mu_B$ and 
and $\mu_{Fe2}$ = 1.86  $\mu_B$, cf. Tab.\ \ref{tab:X}. For the \Fe1 the ferromagnetic coupling
is promoted by the ferromagnetic next neighbors Co at a distance of 2.48 {\AA}, and
the locally increased valence charge leads to an enhancement of Fe moments
(see Slater-Pauling curve\cite{sl37,co}).
The non-magnetic Ga atoms lie at the distance of 2.86 {\AA} to the \Fe1 atoms.
For the \Fe2 atoms the positions of the Co and Ga are exchanged and thus
the non-magnetic Ga atoms take 4 of the 8 next neighbor positions. Thus
it is conceivable that $\mu_{Fe2}$ = 1.86 $\mu_B$ is even below the
usual 2.2$\mu_B$ of the bcc Fe single crystal.\cite{ii13}

In Fig.\ \ref{fig:vol_vs_locmag} (b) we show the distribution of local magnetic moments and Voronoi volumes that we get 
for the magnetic elements in the inverse Heusler GBs. Also here we observe large fluctuations in $\Vat$ for atoms situated at or near a GB plane. Moreover, we see the strong dependence on the local atomic environment in the spread of data points lying in between the two limiting cases for $\muat$ of \Fe1 and \Fe2 in the bulk single crystal, as illustrated by the solid lines. Co atoms at or close to the GB plane as well may have an increased volume compared to the bulk equilibrium and a $\muat$ that on average is larger by 0.2 $\mu_B$ than expected when straining a bulk crystal to obtain the corresponding increase in $\Vat$. Note that most Co and \Fe1 atoms with a reduced $\Vat$ compared to their equilibrium value also have a considerably increased $\muat$. This is caused by the modified atomic neighborhood of those atoms at the GB.

\begin{figure}[t]
\begin{center}
	\includegraphics[clip, trim=2.cm 21.3cm 10.4cm 2.1cm,width=\columnwidth]{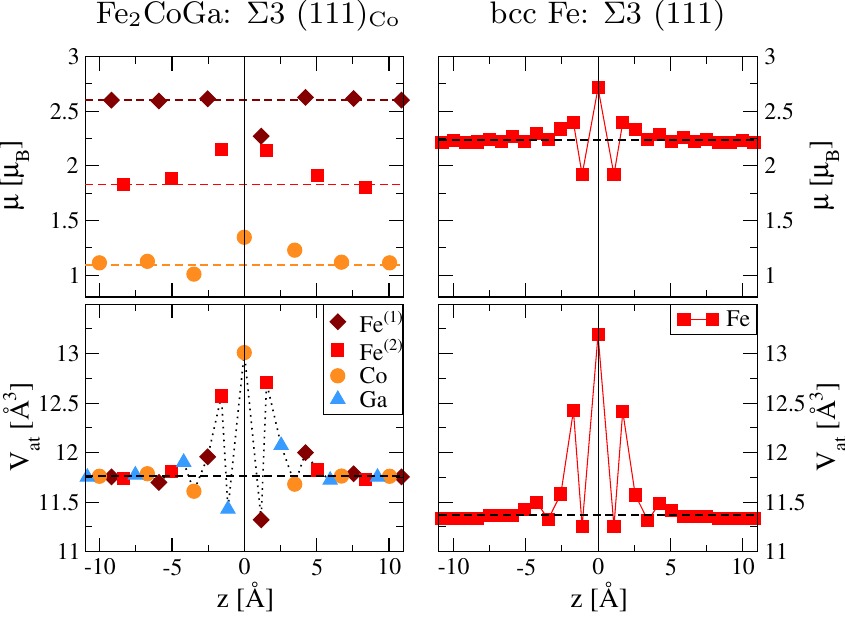}
\end{center} 
\caption{(Color online) Local magnetization (top) and Voronoi volume (bottom) for atoms in the \mbox{$\Sigma 3$ (111)} GB of the inverse Heusler Fe$_2$CoGa (left) 
and compared to bcc Fe (right). \label{fig:magvol_sig3_111} }
\end{figure}

\begin{figure}
\begin{center}
	\includegraphics[clip, trim=2.cm 21.3cm 10.4cm 2.cm,width=\columnwidth]{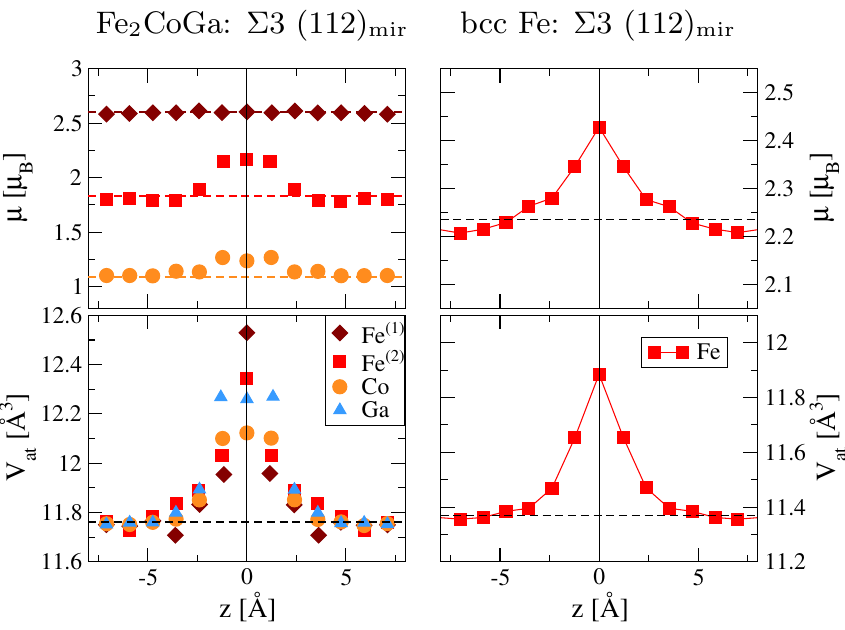}
\end{center} 
\caption{(Color online)  Local magnetization (top) and Voronoi volume (bottom) for atoms in the mirror symmetric \mbox{$\Sigma 3$ (112)$_{\rm mir}$} GB of the inverse Heusler Fe$_2$CoGa  (left) and compared to  bcc Fe (right). \label{fig:magvol_sig3_112mir} }
\end{figure}

\begin{figure}
\begin{center}
	\includegraphics[clip, trim=2.cm 21.3cm 10.4cm 2.cm,width=\columnwidth]{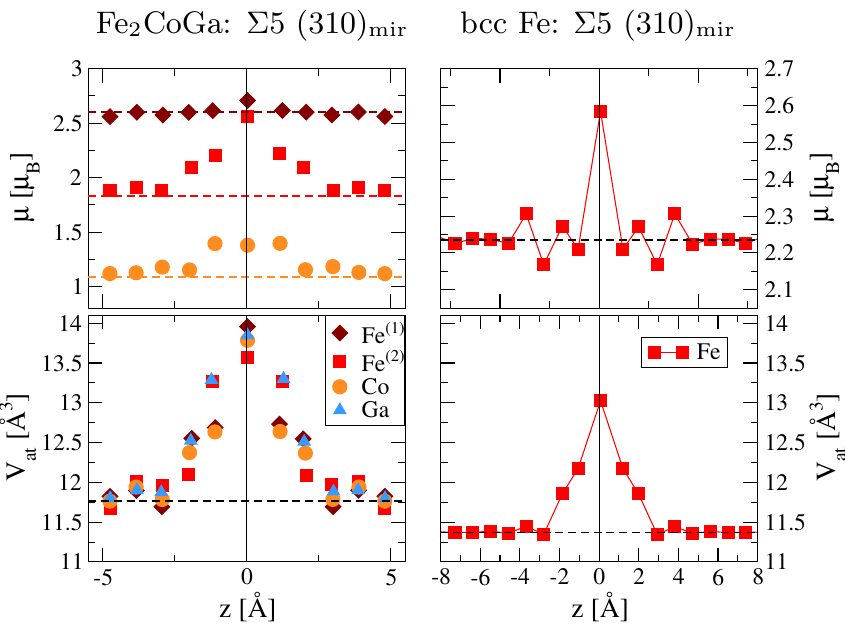}
\end{center} 
\caption{(Color online) Local magnetization (top) and Voronoi volume (bottom) for atoms in the mirror symmetric 
\mbox{$\Sigma 5$ (310)$_{{\rm mir}}$} GB of the inverse Heusler Fe$_2$CoGa (left) and compared to bcc Fe (right). \label{fig:magvol_sig5_310mir} }
\end{figure}

\begin{figure}
\begin{center}
	\includegraphics[clip, trim=2.cm 21.3cm 10.4cm 1.7cm,width=\columnwidth]{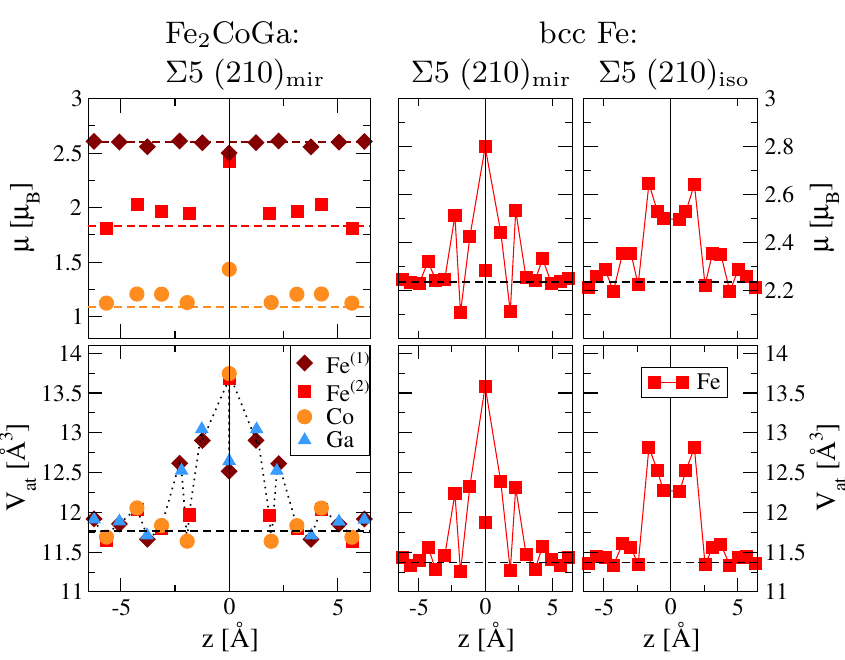}
\end{center} 
\caption{(Color online) Local magnetization (top) and Voronoi volume (bottom) for atoms in the mirror symmetric 
\mbox{$\Sigma 5$ (210)$_{\rm mir}$} GB of the inverse Heusler Fe$_2$CoGa (left) and in bcc Fe (middle). The right panels show the results for the $\Sigma 5$ (210$)_{\rm iso}$ in bcc Fe for comparison. \label{fig:magvol_sig5_210mir} }
\end{figure}

In order to analyze our GB models in more detail, 
we evaluate the spatial distribution of local magnetic moments and Voronoi volumes in the vicinity of the GB plane.
In the following we discuss the results for the cases of $\Sigma 3 (111)$, $\Sigma 3 (112)_{\rm mir}$, $\Sigma 5 (310)_{\rm mir}$, and $\Sigma 5 (210)_{\rm mir}$ which are shown in Figs.\ \ref{fig:magvol_sig3_111}, \ref{fig:magvol_sig3_112mir}, \ref{fig:magvol_sig5_310mir}, and \ref{fig:magvol_sig5_210mir}, respectively. In all four figures, we compare the results for the inverse Heusler GB with the corresponding one for bcc Fe. Local magnetic moments in the top panels and Voronoi volume in the lower panels are plotted as function of absolute distance (in {\AA}) of the respective atom from the GB plane, which is set to $z=0$ in all cases.  

The $\Sigma$3 (111) GB shows a correlated oscillatory distance dependence of $\Vat$ within a range of approx.\ $\pm 4$ {\AA} from the GB plane for both, the inverse Heusler Fe$_2$CoGa and bcc Fe GBs (cf. Fig.\ \ref{fig:magvol_sig3_111}). The magnitude of the volume modulation is very similar in both cases.
As the $\Sigma$3 (111) structure consists of a stacking of mono atomic (111) lattice planes (cf.\ Fig.\ \ref{fig:GBsupercells}) the essential degree of freedom in structural relaxation is a change in layer spacing with a maximum excess widening at the GB plane.
The local magnetic moments also follow oscillatory behavior which is directly correlated with the modulation in $\Vat$.
This is clearly visible for the Fe atoms near the GB in bcc Fe. The increase/decrease of $\muat$ for atoms in the inverse Heusler phase GB is directly coupled to the increase/decrease of $\Vat$, too. One major difference is observed between the two cases: Contrary to the inverse Heusler phase, for bcc Fe the oscillatory behavior averages out to a large extent, and the integrated change in magnetization is comparably low, cf.\ Fig.\ \ref{fig:GB_magn_prop}.

The situation is different for the $\Sigma 3$ (112) GB, shown in Fig.\ \ref{fig:magvol_sig3_112mir}. Here there are no oscillations, instead a peak-like modulation around the GB plane is observed for $\Vat$ and $\muat$. Note that the magnitude of modulation as well as the affected region around the GB is reduced compared to the $\Sigma 3$ (111) but similar for both phases. This is reflected in similar values for the integrated change in magnetization $\xi_M\sim 5\mu_B/$nm$^2$. However, while for bcc Fe there is a one-to-one correspondence between $\Vat$ and $\muat$, the local magnetic moments in the inverse Heusler phase Fe$_2$CoGa now strongly depend on the atomic elements in their modified local atomic environments. This can be seen for the central three \Fe1 atoms that have rather different volumes of 12.53 {\AA}$^3$ and 11.96 {\AA}$^3$ but almost the same $\muat\simeq 2.6\mu_B$ as in the bulk region. The three central \Fe2 atoms have as well an almost identical increase in $\muat$ but different volumes. For the \Fe2 atom at $z=0$, the first neighbor shell is deteriorated and one of the originally four \Fe1 atoms is replaced by a non-magnetic Ga atom. The situation is reversed for the other two \Fe2 atoms which can achieve the same moment enhancement with lower volume by having more Fe atoms in their first neighbor shell. Due to the presence of the GB plane they have in total five neighboring Fe atoms. 

As a third case, we discuss the $\Sigma 5$ (310) which shows larger structural variations at the GB plane. Again, there is a peak-like increase of local volumes at the GB for both considered phases (cf. Fig.\ \ref{fig:magvol_sig5_310mir}). However, in the case of bcc Fe, the local magnetic moment oscillate with a dominant maximum of approx. 2.6 $\mu_B$ for the atoms lying on the GB plane. The effect on the local magnetic moments is more pronounced and extended on a larger length scale in the inverse Heusler Fe$_2$CoGa GB. This can again be traced back to the change in distance and element species of neighboring atoms. As a consequence, the integrated change in magnetization $\xi_M$ for this GB type is considerably lower in bcc Fe than in the inverse Heusler Fe$_2$CoGa (cf.\ Fig.\ \ref{fig:GB_magn_prop}).

The $\Sigma 5$ (210) has the overall largest excess magnetization of all the GB types considered in this work (see Fig.\ \ref{fig:magvol_sig5_210mir}). For Fe$_2$CoGa, mainly the Co and \Fe2 atoms on the GB plane are responsible for this elevated magnetization. Interestingly, the \Fe1 atoms in the neighboring atomic planes do not contribute to this increase, although they have a substantially enlarged $\Vat$. The largest $\xi_M$ value is found for the $\Sigma 5$ (210)$_{\rm mir}$ in bcc Fe (right panel in Fig.\ \ref{fig:magvol_sig5_210mir}). Here the geometric configuration does not lead to the oscillations in $\Vat$ and $\muat$ observed for the other GBs in bcc Fe. Instead we find a broad double peak at the GB in both quantities, which is the origin for the observed extra large excess magnetization. This effect is not seen in the GBs of isoceles type in the inverse Heusler phase (cf. Tab.\ \ref{tab:results:GB}). Here again, the ternary alloy Fe$_2$CoGa provides higher flexibility to react to the structural perturbation of the GB.

\section{Summary and Conclusions}
\label{sec:summary}

We have presented a DFT study of the energetic and magnetic properties of grain boundaries
in the inverse Heusler phase Fe$_2$CoGa and compared the results to corresponding GBs in bcc Fe.
The studied GBs in the inverse Fe$_2$CoGa phase all have comparably lower formation energies than the respective ones in bcc Fe. However, the surface formation energies and the work of separation are rather similar for both phases. For all the considered structures, we find that the presence of GBs leads
to a systematic enhancement of the local magnetic moments in the GB region and thereby to a positive integrated excess magnetization. The fact that Fe$_2$CoGa contains nonmagnetic Ga and thus 25\%  of the atoms are not profiting from a volume increase in the GB region is compensated by Co and \Fe2 atoms
which are mainly responsible for the gain in magnetic moment at the GBs.

The GBs of $\Sigma 3$ (112) type are by far the most stable GBs in both material systems and very similar in terms of their geometric, energetic and magnetic properties.
For the other GB types studied, namely the $\Sigma 3$ (111), $\Sigma 5$ (210), and $\Sigma 5$ (310), pronounced differences in energetic and magnetic properties are observed.
The respective GBs in Fe$_2$CoGa lead to a considerable increase of magnetization at the GB, up to more than twice as much as in bcc Fe, depending on the GB type, while geometrical quantities like GB excess widening or local volume distributions are similar for both phases. We have analyzed this difference in terms of the distribution of magnetic moments and volumes of the atoms in vicinity of the GB. We find that the ternary Fe$_2$CoGa phase yields a higher flexibility in compensating the disturbance of crystal defects by structural relaxation. Thereby it forms GB structures of lower energy accompanied with increased local magnetic moments of the Co and \Fe2 atoms within a distance of a few $\ang$ around the GB plane. 

\acknowledgments 
We thank Dr.\ Georg Krugel for collaborating with us in an early stage of this investigation. This work was funded by a Fraunhofer--Max-Planck cooperation project "HEUSLER - New magnetic materials without rare earth elements". 
Parts of the calculations were performed on the computational resource ForHLR II funded by the Ministry of Science,
Research and the Arts Baden-W\"urttemberg and DFG ("Deutsche Forschungsgemeinschaft").

\bibliography{Heusler}

\end{document}